\documentclass[sigconf]{acmart}
\AtBeginDocument{%
  }

\usepackage{algorithm}
\usepackage{algpseudocode}
\usepackage{complexity}
\usepackage{graphicx}
\usepackage{balance}
\usepackage{stfloats}
\usepackage{comment}
\usepackage{tabularx}
\usepackage{booktabs}
\usepackage{textcomp}
\usepackage{makecell}
\usepackage{xcolor}
\usepackage{pgfplots}

\pgfplotsset{compat=newest}

\copyrightyear{2026}
\acmYear{2026}
\setcopyright{cc}
\setcctype{by-nc-nd}
\acmConference[ISLPED '26]{ACM/IEEE International Symposium on Low Power Electronics and Design}{August 05--07, 2026}{Evanston, IL, USA}
\acmBooktitle{ACM/IEEE International Symposium on Low Power Electronics and Design (ISLPED '26), August 05--07, 2026, Evanston, IL, USA}
\acmDOI{10.1145/3816440.3818544}
\acmISBN{979-8-4007-2748-1/2026/08}

\begin{document}

\title[A Noise-Aware Quantum Resource Allocation Framework]{How Many Shots Does It Take? A Noise-Aware Quantum \\Resource Allocation Framework}
%% --- Author(s) ---
% left empty for now
% \author{Prateek P. Kulkarni, Sumit K. Mandal \\
% Indian Institute of Science (IISc), Bengaluru, India}
\author{Prateek P. Kulkarni and Sumit K. Mandal}
\affiliation{%
  \institution{Indian Institute of Science (IISc), Bengaluru, India}
  \country{}
}
% \affiliation{
%  \institution{PES University}
%  \city{Bengaluru}
%  \country{India}
% }
% \email{pkulkarni2425@gmail.com}

% \author{Sumit K. Mandal}
% \affiliation{
%     \institution{Indian Institute of Science}
%     \city{Bengaluru}
%     \country{India}
% }
% \email{skmandal@iisc.ac.in}


\begin{abstract}
% Any algorithm executing on quantum computers require several numbers of repeated executions (known as shots) to obtain reliable results.
% However, due to limited availability of quantum computers worldwide, each shot is costly.
% The knowledge of exactly required number of shots is extremely important to reduce the cost.
% In this work, we propose a closed form accurate analytical expression for optimal number of shots required for reliable execution of any algorithm on a given quantum computer.
% The proposed analytical expression is a function of algorithm property, fidelity as well as noise of the underlying quantum computer.
% \todo{In addition, we present an optimal shot budget formulation that specifies how a fixed number of measurement shots should be distributed across different points in a quantum circuit to minimize total estimation error. This allocation principle ensures the most efficient use of available quantum resources while maintaining reliability across all measurement observables.}
% We show that our proposed analytical model helps to reduce the shots associated to reliable execution of algorithms on quantum computers by up to 73\% on average compared to current practice.

Any algorithm execution on quantum computers requires several repeated and costly executions (known as shots) to obtain reliable results. In this work, we propose a closed-form accurate analytical expression to determine optimal number of shots required for reliable execution of any algorithm on a quantum computer. We also present a theoretically grounded technique to distribute fixed shot budget across different partitions in a quantum circuit minimizing the total error. Our proposed analytical model helps to reduce the shots associated with reliable execution of quantum algorithms by about 58\% compared to current practice, in turn reducing the energy consumption by upto 62\%. Furthermore, our proposed optimal shot allocation technique across different partitions reduces total error by up to 73\% compared to conventional approaches.

% \ppk{Additionally, we also determine the minimum number of partitions required when circuits exceed hardware depth limits and optimally allocate shots across these partitions, reducing total error by up to 73\% compared to conventional approaches.}

\end{abstract}


% \ccsdesc[300]{Mathematics of computing~Probability and statistics}




% \settopmatter{printacmref=true}
\maketitle

\begin{figure}
    \centering
    \includegraphics[width=1\linewidth]{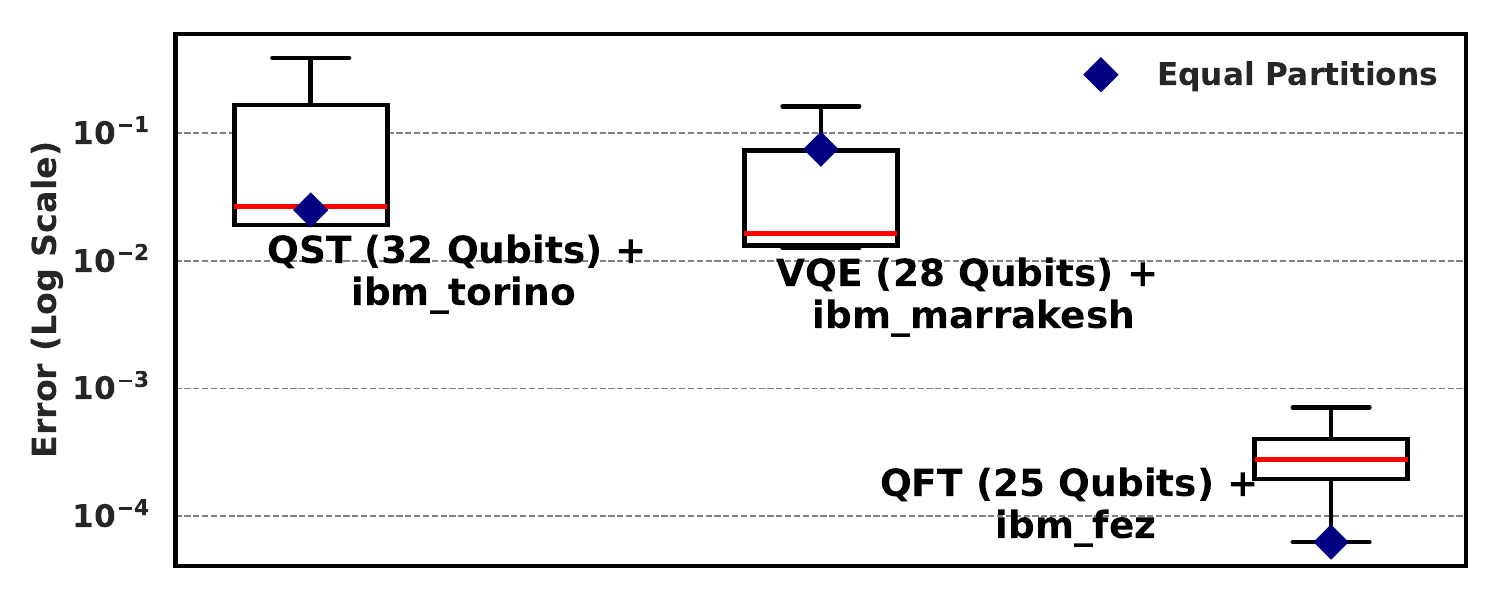}
    %\caption{Distribution of estimation errors for multiple quantum algorithms under fixed shot budget constraints. The box plots demonstrates that error for equal partition shows substantial variation in error across algorithms.}
    \vspace{-8.5mm}
    \caption{
    % Estimation error for 
    % % different 
    % quantum algorithms under a fixed shot budget, with all partitioning strategies. 
    % % with different partitioning strategies. 
    % Equal shot partitioning displays substantial variation in error across algorithms.
    Estimation error for quantum algorithms across all partitioning strategies, under a fixed shot budget. Equal shot partitioning shows substantial variation across algorithms.
    }
    \label{fig:error_partitioned}
    \vspace{-6.5mm}
\end{figure}

\section{Introduction}
\label{sec:introduction}

% Quantum computing holds the promise of solving certain problems exponentially faster than classical computers, with the complexity class $\BQP$ capturing most of the algorithms believed to achieve this advantage.

The practical realization of any algorithm on near-term quantum computers is hindered by noise, limited gate fidelities, and the need for repeated executions (known as shots) to obtain reliable results. These challenges raise fundamental questions about the minimal requirements for quantum computers to achieve reliable and efficient computation. In particular, understanding the trade-offs between the number of shots, fidelity, and success probability is crucial for assessing the feasibility of near-term quantum algorithms and for identifying regimes where quantum computers can outperform classical systems. 

% Quantum computing holds the promise of solving certain problems exponentially faster than classical computers, with the complexity class capturing most of the algorithms believed to achieve this advantage. However, the practical realization of these algorithms on near-term quantum computers is hindered by noise, limited gate fidelities, and the need for repeated executions (or, shots) to obtain reliable results. These challenges raise fundamental questions about the minimal requirements for quantum computers to achieve reliable and efficient computation. In particular, understanding the trade-offs between the number of shots, fidelity, and success probability is crucial for assessing the feasibility of near-term quantum algorithms and for identifying regimes where quantum computers can outperform classical systems.

Currently, researchers rely on two primary approaches to evaluate quantum algorithms -- simulation on classical computers or execution on real quantum hardware.
However, simulations often make impractical assumptions; such as perfect noise models or infinite coherence times which may not hold in practice, leading to unreliable predictions. On the other hand, real quantum computers are not yet readily available at scale, and access to sufficiently powerful hardware remains limited.
Due to limited availability of quantum computers, each shot performed on a quantum computer is expensive.
Therefore, it is important to limit the number of shots while achieving reliable execution.
% \todo{Mention that we are modeling the number of shot as a function of p(s).
% After that motivate why do we need to partition the algorithm and why shots at each partition matter.
% Then mention how are we addressing the challenge.}
% \ppk{
% In this work, we try to resolve this challenge as follows. We first derive an expression for the number of shots as a function of probability of success. This relation allows explicit calculation of exact shots required to amplify a computation to desirable probability, given the knowledge of hardware parameters.
% Compared to current practice, we show an average reduction of about 58\% in the number of shots.}
% 
To this end, we first derive a closed-form analytical expression for the number of shots as a function of probability of success. This relation allows explicit calculation of exact number of shots required to achieve desired reliability while executing an algorithm on quantum computer, given the knowledge of hardware parameters.

Besides, given a fixed shot budget, a quantum computer allows computations of only up to certain depth of an algorithm.
If the algorithm has a gate depth beyond that, it becomes necessary to partition it into smaller sub-circuits with depths that can be accommodated by the hardware.
One can execute the algorithm on a larger quantum computer but that requires higher cost subject to availability of the quantum computer.  
% Another straightforward option could be to migrate to a larger quantum computer, but that creates another set of problems. The noise profile changes significantly. And, since the topology changes, any topology-based optimization and qubit mapping is lost. Moreover, larger the quantum computer, the more expensive the cost is per shot. Thus, a more economical and efficient route is to partition.
It is a non-trivial task to distribute the total number of shots among all the partitions of the algorithm.
Naively allocating equal number of shots across all sub-circuits would be sub-optimal, as it would disregard noise variances, circuit depths and qubit resources specific to a single sub-circuit, leading to significant error accumulation over multiple shots.
Figure~\ref{fig:error_partitioned} shows the range of errors when different algorithms are executed on different quantum computer under a fixed shot budget.
In this case, we sweep all the possible number of shots for different partition of the algorithm.
% \todo{For example, we consider QFT algorithm to be executed on IBM Marrakesh consisting of 156 qubits. However, QFT requires 300 qubits (base depth of 60, which expands to 300 after applying five rounds of error-correction, exceeding the hardware depth limit of 285 (for ibm\_marrakesh) and therefore requiring at least \texttt{ceil}(300/60) = 5 partitions) in total and only 156 qubits are available. Therefore, we need to partition QFT in 5 ways.}
For example, we consider QFT algorithm to be executed on IBM Marrakesh which supports a maximum depth of 285 (estimated using our analytical model (Eq.~\ref{eq:d_max}) based on device calibration parameters (e.g., $T_1$, $T_2$, gate duration, and fidelity) obtained from the IBM Quantum Platform~\cite{ibmComputeResources}).
However, QFT algorithm requires a depth of 1200 including error correction.
Therefore, QFT needs to be partitioned by at least $\lceil 1200/285 \rceil = 5$ ways.
Let us also assume that the total number of shots we need to perform for reliable execution is 8000. Now, we need to distribute 8000 shots across 5 ways. We consider all possible distributions and execute QFT accordingly. Upon execution, we obtain the range of errors which is shown in Figure \ref{fig:error_partitioned}.
%the figure. 
It is seen that the error of execution is a strong function of the distribution of shots. Moreover, the error for equal distribution of shots among all partitions can result in different error. For example, the equal distribution results in very less error for QFT on IBM Marrakesh but results in very high error for VQE on IBM Fez.
To this end, we formulate an optimization problem to find the best shot allocation strategy that guarantees minimum total errors over the full circuit.
Specifically, we minimize the sum of squared measurement errors across partitions under a fixed shot budget using Lagrange multipliers, and derive an exact allocation formula where shots are distributed proportionally to each partition's noise variance, so that partitions with higher noise are assigned more measurements.
On average, our proposed technique reduces the total error rate by about 53\% and total number of shots by about 59\% across multiple quantum algorithms executing on different quantum computers.
The major contribution of this work are as follows:
\begin{itemize}
    \item A closed-form expression relating shots $s$ to success probability $P(s)$ that incorporates hardware noise parameters including decoherence times and gate errors.

    % \item We establish an exact analytical expression for maximum executable circuit depth $d_{\text{max}}$ based on fundamental hardware capabilities and target reliability.

    % \item \ppk{We derive the minimum number of partitions $m_{\text{min}}$ required when circuit depth exceeds hardware limits, enabling systematic circuit decomposition.}

    \item An optimal shot allocation strategy that distributes shots amongst all partitions, proportionally to noise variances, minimizing total error under fixed budget constraints.

    \item Extensive experimental evaluation executing various algorithm on real quantum computer with our proposed allocation strategy showing 63\% improvement in estimation error compared to existing technique, resulting in reduction in energy consumption by 59\% on average.
\end{itemize}

% \todo{add part about depth about error correction}
% This work is motivated by the need to rigorously characterize the trade-offs in quantum computation while accounting for these practical constraints, providing a framework to assess the true limitations of near-term quantum algorithms in under realistic noise and resource constraints.

% \begin{figure}
%     \centering
%     \includegraphics[width=1\linewidth]{images/costvsshots.png}
%     \caption{\todo{Expensive cost scaling with shots, and not sure how many shots is good enough.}}
%     \label{fig:cost}
% \end{figure}

\vspace{-3mm}
\section{Related Work}
\label{sec:related-work}
\vspace{-1mm}

The challenge of determining the number of measurement shots is central to executing algorithms reliably on near-term quantum computers. Recent studies have highlighted that excessive measurement repetition is a major cost driver, underscoring the need for efficient shot allocation~\citep{Kessler2023, preskill2018quantum}. While some work provide shot analyses for specific algorithms like Grover's search~\citep{Kessler2023}, a general, closed-form relationship between shots, success probability, and hardware noise has been missing.

Adaptive methods have been proposed to allocate shots dynamically. For example, some studies redistribute shots among terms in a quantum algorithm to reduce total variance~\citep{zhu2024vqe, phalak2023qmlshots}, while others use machine learning to adjust shot counts during computation \citep{liang2024ai}. These methods show practical gains but are often based on heuristics rather than analytical noise models.

% Connecting hardware noise to algorithmic performance is another active area.
Several work have explored the relation between hardware noise and algorithmic performance.
A recent work has derived how hardware parameters like decoherence times and gate errors contribute to measurement variance~\citep{seksaria2026estimatingshotsvariancenoisy}. Other approaches use overall hardware metrics, like Quantum Volume, to estimate device capability, or employ error mitigation techniques to improve result quality—though often at a high cost in additional runtime and shots~\citep{temme2017errormitigation, mezher2025bqp}.

For deep circuits that exceed hardware limits, partitioning is a common strategy. A variety of methods have been developed for this purpose, including graph-based algorithms that cut circuits along qubit connections~\citep{martinez2019hypergraph, Brandhofer2023partition} and scalable hybrid frameworks like CutQC~\citep{tang2021cutqc}, which decompose large circuits into smaller, executable subcircuits. While these approaches effectively manage depth, connectivity, and qubit mapping constraints, they treat the quantum execution of each partition as a separate, isolated step, without optimizing the allocation of measurement resources across them. Theoretical studies have also explored the limits of shallow circuits~\citep{bravyi2018quantum}, but none provide explicit formulae for the maximum executable depth or the optimal number of partitions based on fundamental hardware parameters.
Barron et al.~\cite{barron2024provable} derives statistical bounds for estimating noise-free expectation values from noisy quantum measurements, showing that sufficiently many shots can mitigate measurement noise. In contrast, our work derives hardware-aware closed-form expressions for the optimal number of shots and their allocation across circuit partitions, enabling resource-efficient execution of quantum circuits on noisy devices.

In summary, prior work has addressed fragments of the problem (shot allocation, noise modeling, and circuit partitioning), but have not provided a unified framework. Our work closes this gap by deriving exact closed-form expressions for the required shots, the maximum circuit depth a quantum machine can support, and the optimal distribution of a shot budget across circuit partitions to minimize total error of execution.

\vspace{-3.5mm}
\section{Background}
\label{sec:preliminaries}
\vspace{-1mm}
\noindent{\textbf{Quantum Circuit Model:}} A quantum circuit $\mathcal{C}(n, d)$ operates on $n$ qubits with depth $d$, representing the path with most gates in the circuit. Circuit depth fundamentally limits feasible computations on current quantum hardware. Quantum states degrade continuously during execution due to environmental noise -- longer computations accumulate more errors. This creates a practical threshold: circuits exceeding a certain depth produce results too corrupted to be useful. Estimating this threshold is challenging as error accumulation depends on gate types, qubit connectivity, and runtime noise that varies across executions. Additionally, most available depth is consumed by error correction circuitry, leaving limited budget for the actual algorithm~\cite{gidney2021factor}. Throughout this work, when we refer to circuit depth $d$, we assume it \textit{includes both the algorithmic operations and the error correction overhead}, which varies depending on the choice of error correction code family. The circuit execution time is $t_{\text{circ}} \approx d \cdot t_g$, where $t_g$ is the average gate duration. This approximation works well because quantum circuits typically employ a relatively uniform mix of one- and two-qubit gates with similar execution times, and modern compilers balance operations across layers to minimize idle time. This averaged
% average-case 
model is widely used in quantum computing for resource estimation and scheduling \cite{nielsen2010quantum}.

A \textit{shot} is one complete circuit execution. Because quantum computers produce probabilistic outputs, multiple shots are required to determine results through statistical aggregation.

\noindent{\textbf{Hardware Reliability Parameters}.} 
Hardware reliability is characterized by several key parameters. The \textit{maximum qubit fidelity} ($F_{\max}$) represents the highest measured reliability across all qubits, typically derived from readout calibration data as $F_{\max} = 1 - p_{\text{readout}}$, where $p_{\text{readout}}$ is the readout assignment error rate. Readout assignment error refers to 
% inaccuracies in the measurement process that occur at the end of quantum computation, that is, it quantifies 
the average probability of incorrectly measuring the state of a qubit. Quantum hardware vendors report $F_{\max}$ through standardized benchmarking protocols, providing a best-case baseline for system quality~\cite{readout-error}. Values range from 0 (completely unreliable) to 1 (perfect). Note that $F_\text{max}$ captures only the measurement; decoherence and gate noise enter via 
Equation \ref{eq:noise}.
% \todo{Quantum states degrade over time through two mechanisms: \textit{energy relaxation time} $T_1$ measures how long qubits retain their energy state, while \textit{dephasing time} $T_2$ measures how long superposition states remain coherent. Typically, $T_2 \leq 2T_1$. The \textit{gate error probability} $p_g$ provides the failure probability per qubit operation.}

% \noindent{\textbf{Success Probability:}} 
% The \textit{success probability} $P(s)$ quantifies the likelihood that statistical aggregation over $s$ shots produces the correct algorithmic output. Since quantum algorithms are inherently probabilistic, achieving reliable computation requires $P(s)$ to meet a target confidence threshold (usually \textgreater 95\%). However, determining how many shots are needed to reach this threshold is challenging as it depends on circuit depth, hardware fidelity, and accumulated noise, which interact in complex ways. Current practice often relies on empirical tuning or conservative over-provisioning of shots without principled analysis.

\vspace{-4mm}
\section{Methodology}

\begin{figure*}[t]
    \centering
    \includegraphics[width=1\linewidth]{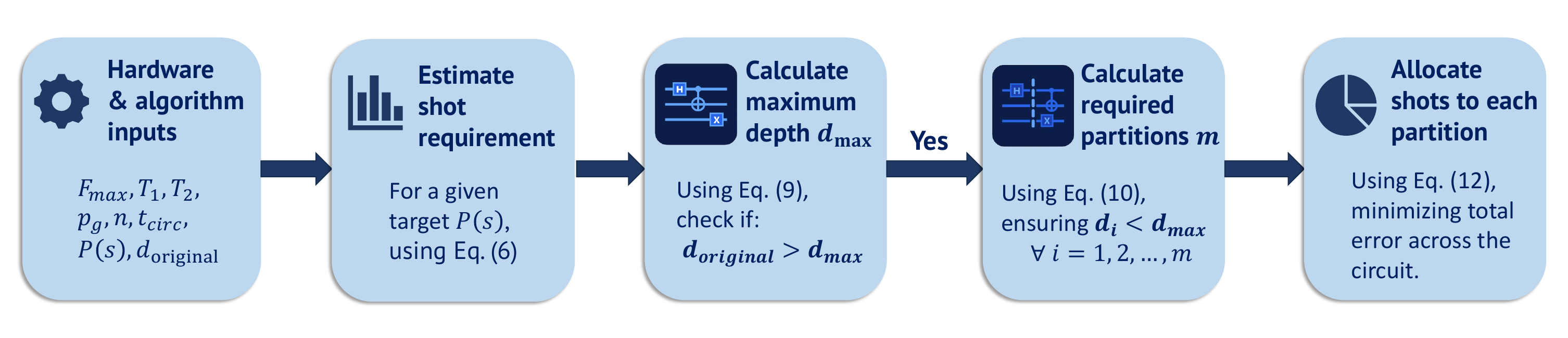}
    \vspace{-11mm}
    \caption{Overview of the Proposed Framework. The workflow begins with hardware parameters and algorithm specifications, estimates total shot requirements, computes maximum executable depth, partitions circuits exceeding depth limits, and optimally distributes shots across partitions to minimize error while achieving target success probability.}
    \vspace{-5mm}
    \label{fig:framework_overview}
\end{figure*}

%\subsection{Overview of the Proposed Framework}

% \noindent\textbf{Overview of the Proposed Framework:}
Our framework provides a systematic approach to optimize number of shots required to execute a given quantum circuit under hardware constraints, as illustrated in Figure~\ref{fig:framework_overview}.
% The process begins with 
The hardware parameters ($F_{max}$, $T_1$, $T_2$, $p_g$) and algorithm specifications ($n$, $t_{circ}$, $P_s$, $d_{original}$) are inputs, and proceeds through a series of optimization stages to determine the optimal execution strategy.

The first stage estimates the total number of measurement shots required to achieve the target success probability $P_s$ using Equation~\ref{eq:shot}. This estimation accounts for the statistical uncertainties inherent in quantum measurements and establishes the computational budget for the entire circuit execution.
% Understanding the shot requirement is crucial as it directly impacts both the runtime and the achievable accuracy of the quantum algorithm.
% 
With the shot budget established, the next step is to compute the maximum executable circuit depth $d_{max}$ via Equation~\ref{eq:d_max}, leveraging the decoherence times and gate fidelities of the quantum hardware. This threshold represents the deepest circuit that can maintain acceptable fidelity on the given quantum hardware. The maximum depth acts as a fundamental constraint which is required in the next step.
% that shapes all subsequent optimization decisions.
% 
As shown in the figure, the framework next evaluates whether the original circuit depth $d_{original}$ exceeds the hardware-imposed limit $d_{max}$. If the circuit is too deep to execute directly, we calculate the required number of partitions $m$ using Equation~\ref{eq:optim_m}. This decomposition ensures that the depth $d_i$ of each partition satisfies $d_i < d_{max}$ for all $i = 1, 2, \ldots, m$, thereby enabling execution of deep circuits that would otherwise be infeasible on near-term quantum devices.

Finally, we distribute the total shots
% (obtained in the second step)
across all partitions using Equation~\ref{eq:optimal}, minimizing the total error while maintaining a target success probability $P_s$.
% This allocation strategy 
This ensures that computational resources are optimally distributed across partitions, balancing the error contributions from different circuit segments to maximize fidelity. The result is a complete execution plan that respects hardware limitations while achieving the desired accuracy guarantees.
% for the computed results.

\vspace{-3mm}
\subsection{Modeling Number of Shots}\label{sec:main-result}

% Given

% \begin{theorem}[Probability of Success Bound] \label{thm:success-bound}
% Let $s \in \mathbb{N}$ denote the number of shots, and $F_{\text{max}} \in [0, 1]$ denote the maximum fidelity of the quantum operation. The probability of success $P(s)$ is bounded as follows:
% \begin{equation}
% P(s) = \frac{1}{2} \left[ 1 + \text{erf} \left( \frac{s F_{\text{max}} - s/2}{\sqrt{2 (s F_{\text{max}} (1 - F_{\text{max}}) + \sigma_\text{noise}^2})} \right) \right],
% \end{equation}
% where $\text{erf}(\cdot)$ denotes the error function and $\sigma_\text{noise}^2$ represents the additional noise contribution arising from quantum communication between qubits, including entanglement degradation, crosstalk, and channel imperfections.
% \end{theorem}

% \begin{proof}

The successive shots, although influenced by noise, are unaffected by one another. Thus one can model the success probability of these shots using a Bernoulli random variable~\cite{Kessler2023}. We define $X_i$ as the outcome of the $i$-th shot, where $X_i \in \{0, 1\}$ is a Bernoulli random variable with success probability $F_{\text{max}}$.
Here, $F_{\text{max}}$ is the maximum fidelity of the quantum computer under consideration.
Then, we have $\mathbb{P}(X_i = 1) = F_{\text{max}}$ and $\mathbb{P}(X_i = 0) = 1 - F_{\text{max}}$. Let $S_n = \sum_{i=1}^n X_i$ denotes the total number of successes after $n$ shots. Thus, $S_n$ follows a binomial distribution:
\vspace{-2mm}
\begin{equation}
\mathbb{P}(S_n = k) = \binom{n}{k} (F_{\text{max}})^k (1 - F_{\text{max}})^{n-k}, \quad k = 0, 1, \ldots, n.
\end{equation}
% 
% \vspace{-1mm}
For $s$ shots, the mean and variance of $S_n$ are $\mu = s F_{\text{max}}$ and $\sigma^2 = s F_{\text{max}} (1 - F_{\text{max}})$, respectively. As $s$ grows large, we can use the Central Limit Theorem to approximate the distribution of $S_n$ by a normal distribution $S_n \sim \mathcal{N}(\mu, \sigma^2 + \sigma_\text{noise}^2)$,
% 
% \begin{equation} \label{eq:normal_dist}
% S_n \sim \mathcal{N}(\mu, \sigma^2 + \sigma_\text{noise}^2).
% \end{equation}
% 
where $\sigma^2$ represents the intrinsic variance due to the probabilistic nature of quantum measurement, and $\sigma_\text{noise}^2$ corresponds to the additional variance introduced by hardware imperfections and decoherence effects, respectively. We define the cumulative success probability as $P(s) = \mathbb{P}(S_n \geq s/2)$. The success probability captures the likelihood of achieving a majority of successes in $n$ trials.
This definition is rooted in majority-vote criteria often used in fault-tolerant quantum computing, such as in Shor~\cite{shor1994algorithms} and Aharonov \& Ben-Or~\cite{aharonov1997fault}. Note that $P(s)$ characterizes readout-level
reliability rather than algorithm-specific success; since all 
quantum measurements yield binary readout outcomes, the 
majority-vote criterion applies uniformly across algorithms.
Therefore, $P(s)$ can be computed as the cumulative distribution function of $S_n$:
\vspace{-2mm}
\begin{equation}
P(s) = \int_{s/2}^{\infty} \frac{1}{\sqrt{2 \pi (\sigma^2 + \sigma_\text{noise}^2)}} \exp\left(-\frac{(x - \mu)^2}{2 (\sigma^2 + \sigma_\text{noise}^2)}\right) dx
\end{equation}
Standardizing the variable $x$ by substituting $z = \frac{x - \mu}{\sqrt{\sigma^2 + \sigma_\text{noise}^2}}$, with $dx = \sqrt{\sigma^2 + \sigma_\text{noise}^2} dz$, gives:
\vspace{-2mm}
\begin{equation}\label{eq:prob}
P(s) = \int_{z_0}^{\infty} \frac{1}{\sqrt{2 \pi}} \exp\left(-\frac{z^2}{2}\right) dz, \quad z_0 = \frac{s/2 - \mu}{\sqrt{\sigma^2 + \sigma_\text{noise}^2}}
\end{equation}
Substituting $\mu = s F_{\text{max}}$ and $\sigma = \sqrt{s F_{\text{max}} (1 - F_{\text{max}})}$, we find:
\vspace{-2mm}
\begin{equation}
z_0 = \frac{s/2 - s F_{\text{max}}}{\sqrt{s F_{\text{max}} (1 - F_{\text{max}}) + \sigma_\text{noise}^2}}
\end{equation}
The right hand side of Equation~\ref{eq:prob} corresponds to the complementary error function $\text{erfc}(z_0)$, so we can express $P(s)$ as $\text{erfc}(-z_0)/2$. Using the relationship $\text{erfc}(-z) = 1 + \text{erf}(z)$, we obtain $P(s) = \frac{1}{2} \left[ 1 + \text{erf}(z_0) \right]$.
% 
% \begin{equation}
% P(s) = \frac{1}{2} \left[ 1 + \text{erf}(z_0) \right]
% \end{equation}
% 
Substituting $z_0$ back into the expression, we have:
\vspace{-2mm}
\begin{equation}\label{eq:p(s)}
P(s) = \frac{1}{2} \left[ 1 + \text{erf}\left( \frac{s F_{\text{max}} - s/2}{\sqrt{2 (s F_{\text{max}} (1 - F_{\text{max}}) + \sigma_\text{noise}^2})} \right) \right]
\end{equation}
% 
% Equation \ref{eq:p(s)} estimates the probability of success exactly after $s$ shots.
Equation~\ref{eq:p(s)} estimates the probability that an algorithm can be reliably executed on a quantum computer exactly after $s$ shots.
Solving for $s$, we obtain:
\vspace{-2mm}
\begin{equation}\label{eq:shot}
    s = \frac{[\text{erf}^{-1}(2P(s) - 1)]^2\left(2(F_{\max}(1 - F_{\max})) + 2\sigma^2_{\text{noise}}\right)}{(F_{\max} - 1/2)^2}
\end{equation}
Equation~\ref{eq:shot} estimates exactly how many shots are needed to reach the desirable probability of success for a quantum algorithm, given the maximum fidelity that can be achieved ($F_\text{max}$) and the noise characteristics ($\sigma_\text{noise}^2$), of the underlying quantum hardware. 
% \subsection{Connecting Noise Characteristics with Shot Count}\label{subsec:result}
Physical coherence parameters controlling energy and phase decays
$(T_1, T_2)$, gate error probability $p_g$, and circuit time $t_{\text{circ}}$ are the major contributors to $\sigma_\text{noise}^2$ in Equation~\ref{eq:shot}. We use the following noise variance relation from~\cite{seksaria2026estimatingshotsvariancenoisy}, which models decoherence and gate error contributions in quantum circuits, as a function of shot counts:
\vspace{-2mm}
\begin{equation}\label{eq:noise}
\begin{split}
\sigma_\text{noise}^2 = \frac{n_q}{2 s} \Bigg(& 2
- \frac{(t_{\text{circ}}/T_1)^2}{e^{t_{\text{circ}}/T_1}}
- \frac{(t_{\text{circ}}/T_2)^2}{e^{t_{\text{circ}}/T_2}} \\[2pt]
& - \frac{(t_{\text{circ}}/T_1 + 1)^2}{e^{2 t_{\text{circ}}/T_1}}
- \frac{(t_{\text{circ}}/T_2 + 1)^2}{e^{2 t_{\text{circ}}/T_2}}
+ \frac{p_g^2}{2} \Bigg)
\end{split}
\end{equation}
Substituting this into Equation~\eqref{eq:shot} and solving for $s$ (taking the positive, physical root) we obtain:
% \begin{align*}
% s &= \frac{[\operatorname{erf}^{-1}(2P(s)-1)]^2}{(F_{\max}-\tfrac{1}{2})^2}
% \left( 2F_{\max}(1-F_{\max}) + \frac{n_q}{s} D \right), \\[4pt]
% \Rightarrow \;& (F_{\max}-\tfrac{1}{2})^2 s^2 
% - 2[\operatorname{erf}^{-1}(2P(s)-1)]^2 F_{\max}(1-F_{\max})\, s \\[-1pt]
% &\qquad - [\operatorname{erf}^{-1}(2P(s)-1)]^2 n_q D = 0,
% \end{align*}
% where
% \begin{align*}
% D = 2 &- \frac{(t_{\text{circ}}/T_1)^2}{e^{t_{\text{circ}}/T_1}}
% - \frac{(t_{\text{circ}}/T_2)^2}{e^{t_{\text{circ}}/T_2}} \\[2pt]
% &- \frac{(t_{\text{circ}}/T_1 + 1)^2}{e^{2 t_{\text{circ}}/T_1}} 
% - \frac{(t_{\text{circ}}/T_2 + 1)^2}{e^{2 t_{\text{circ}}/T_2}} 
% + \frac{p_g^2}{2}.
% \end{align*}
% Solving this quadratic in $s$ (and taking the positive, physical root) yields
\vspace{-2.5mm}
\begin{equation}\label{eq:sigma_noise}
\begin{split}
s &= \frac{[\operatorname{erf}^{-1}(2P(s)-1)]^2}{2(F_{\max}-\tfrac{1}{2})^2}
\Bigg( 2F_{\max}(1-F_{\max}) \\[2pt]
& \quad + \sqrt{ \big[2F_{\max}(1-F_{\max})\big]^2
+ \frac{4(F_{\max}-\tfrac{1}{2})^2 \cdot n_q D}{[\operatorname{erf}^{-1}(2P(s)-1)]^2} } \;\Bigg)
\end{split}
\end{equation}
where
\vspace{-2mm}
% \begin{align*}
% D = 2 &- \frac{(t_{\text{circ}}/T_1)^2}{e^{t_{\text{circ}}/T_1}}
% - \frac{(t_{\text{circ}}/T_2)^2}{e^{t_{\text{circ}}/T_2}} \\[2pt]
% &- \frac{(t_{\text{circ}}/T_1 + 1)^2}{e^{2 t_{\text{circ}}/T_1}} 
% - \frac{(t_{\text{circ}}/T_2 + 1)^2}{e^{2 t_{\text{circ}}/T_2}} 
% + \frac{p_g^2}{2}
% \end{align*}
\begin{equation*}
D = 2 + \frac{p_g^2}{2} - \sum_{i=1}^{2}\left[\frac{\alpha_i^2}{e^{\alpha_i}} 
+ \frac{(\alpha_i+1)^2}{e^{2\alpha_i}}\right], \quad \alpha_i \triangleq \frac{t_{\text{circ}}}{T_i}
\end{equation*}
% 
% Given hardware parameters, this expression makes Eq. \ref{eq:shot} fully computable. 
Thus we obtain a closed-form relation between number of shots required for reliable execution of an algorithm as a function of quantum hardware parameters.

\vspace{-3mm}
\subsection{Modeling Minimum Number of Partitions}

When the depth of a quantum circuit exceeds the computational limits of the hardware, a common strategy is to partition the circuit into smaller, executable sub-circuits. However, predicting this maximum feasible depth \textit{a priori} is challenging, often leading to inefficient trial-and-error in current experimental practice. The uncertainty results in significant inefficiencies in resource allocation and computation time.

In this section, we leverage the results from Section~\ref{sec:main-result} to establish an exact relationship for the maximum circuit depth.
Furthermore, we derive an expression of how many sub-circuits a given circuit should be partitioned into given a total shot budget.
The expressions are derived from the fundamental hardware parameters, such as qubit coherence time \(T_2\) and gate time \(t_g\) as well as algorithm-specific requirements, including the number of qubits \(n_q\), the target success probability \(P(s)\), and the maximum achievable fidelity \(F_{\text{max}}\).
% We begin with the noise variance model introduced earlier:
% 
% \begin{align*}
% \sigma_\text{noise}^2 = \frac{n}{2 s} \Bigg(& 2 
% - \frac{(t_{\text{circ}}/T_1)^2}{e^{t_{\text{circ}}/T_1}} 
% - \frac{(t_{\text{circ}}/T_2)^2}{e^{t_{\text{circ}}/T_2}} \\[2pt]
% & - \frac{(t_{\text{circ}}/T_1 + 1)^2}{e^{2 t_{\text{circ}}/T_1}} 
% - \frac{(t_{\text{circ}}/T_2 + 1)^2}{e^{2 t_{\text{circ}}/T_2}} 
% + \frac{p_g^2}{2} \Bigg)
% \end{align*}
% 
% To connect this noise to circuit depth,
We first express the circuit execution time, \(t_{\text{circ}}\), in terms of fundamental operations.
We model \(t_{\text{circ}}\) as the product of the circuit depth \(d\) and a characteristic gate time \(t_g\). This approximation, \(t_{\text{circ}} = d \cdot t_g\), is valid for circuits composed of sequential gates or when \(t_g\) represents an effective average gate time across both single- and two-qubit operations. While a more nuanced model could account for different gate types, this mean-gate-time approximation is sufficient for our scaling analysis.

% We now focus on the regime where the circuit is executed well within the coherence limit, \(t_{\text{circ}} \ll T_2\). In this regime, 
Suppressing the higher-order terms in the exponential expression of the noise model, Eq.~\ref{eq:noise} can be simplified 
% leading 
to the approximation $\sigma^2_{\text{noise}} \approx \frac{n_q}{2s} \cdot \left(\frac{d \cdot t_g}{T_2}\right)^2$.
%
% \begin{equation}
% \sigma^2_{\text{noise}} \approx \frac{n_q}{2s} \cdot \left(\frac{d \cdot t_g}{T_2}\right)^2
% \end{equation}
% 
Here the noise variance is expressed as a function of the depth \(d\), the shot count \(s\), and the intrinsic hardware ratio \(T_2/t_g\). Our goal is to determine the maximum depth \(d_{\text{max}}\) for which a quantum algorithm still meets its required reliability threshold, defined by a minimum success probability \(P(s)\).
% We therefore turn to the expression for \(P(s)\) previously derived in Equation~\ref{eq:p(s)}, which is a function of both the binomial variance and the noise variance \(\sigma^2_{\text{noise}}\).
By substituting depth-dependent model of noise, into the expression for \(P(s)\) shown in Equation~\ref{eq:p(s)}, we solve for the depth \(d\) that satisfies the probability constraint. Solving for \(d\) yields the maximum achievable depth:
\vspace{-2mm}
\begin{equation} \label{eq:d_max}
\begin{split}
d_{\text{max}} \approx \frac{T_2}{t_g} \sqrt{\frac{2s}{n_q} \left(\frac{[s(F_{\text{max}}-1/2)]^2}{2[\operatorname{erf}^{-1}(2P(s)-1)]^2} - sF_{\text{max}}(1-F_{\text{max}})\right)}
\end{split}
\end{equation}
% 

% \noindent For a large number of shots \(s\) and a fixed target probability \(P(s)\), the dominant scaling behavior simplifies to:
% %
% \[
% d_{\text{max}} \propto \frac{T_2}{t_g} \cdot \frac{1}{\sqrt{\ln s}}.
% \]
% %
The above relation indicates that the maximum feasible depth increases linearly with the coherence-to-gate-time ratio but only logarithmically with the number of shots, highlighting a fundamental trade-off between circuit complexity and statistical reliability. 

We now leverage the expression for $d_{\text{max}}$ derived above to determine the number of circuit partitions required to execute deep circuits within hardware constraints.
Let $d_{\text{original}}$ be the total depth of the original, unpartitioned circuit. For any partitioning scheme into $m$ sub-circuits with depths $d_1, d_2, \ldots, d_m$, the fundamental execution constraint requires that \emph{every} partition must satisfy the hardware depth limit $\max\{d_1, d_2, \ldots, d_m\} \le d_{\text{max}}$.
% 
% \begin{equation} \label{eq:upper_bound}
% \max\{d_1, d_2, \ldots, d_m\} \le d_{\text{max}}
% \end{equation}
% 
The optimal number of partitions minimizes $m$ while satisfying this constraint. In the best case, partitions can be perfectly balanced, but in general, circuit structure may impose dependencies that create uneven partitions. However, for any valid partitioning, the maximum partition depth must be at least $\max\{d_1, d_2, \ldots, d_m\} \ge d_{\text{original}}/{m}$.
% 
% \begin{equation} \label{eq:lower_bound}
% \max\{d_1, d_2, \ldots, d_m\} \ge \frac{d_{\text{original}}}{m}
% \end{equation}
% 
This relation follows from the pigeonhole principle -- if all partitions were shallower than $d_{\text{original}}/m$, the total depth would be less than $d_{\text{original}}$.
Combining the inequalities
% Equation~\ref{eq:upper_bound} and Equation~\ref{eq:lower_bound} 
we obtain the necessary condition ${d_{\text{original}}}/{m} \le d_{\text{max}}$.
% 
% \begin{equation}
% \frac{d_{\text{original}}}{m} \le d_{\text{max}}
% \end{equation}
% 
Solving for the minimum number of partitions, we obtain $m \ge {d_{\text{original}}}/{d_{\text{max}}} \implies
m_{\text{min}} = \left\lceil {d_{\text{original}}}/{d_{\text{max}}} \right\rceil $.
Substituting the expression for $d_{\text{max}}$ from Equation~\ref{eq:d_max} and simplifying, we obtain:
% \begin{equation}
% m_{\text{min}} = \left\lceil \frac{d_{\text{original}}}{\frac{T_2}{t_g} \cdot \sqrt{\frac{2s}{n_q} \cdot \left(\frac{[s(F_{\text{max}}-1/2)]^2}{2[\text{erf}^{-1}(2P(s)-1)]^2} - sF_{\text{max}}(1-F_{\text{max}})\right)}} \right\rceil
% \end{equation}
% \noindent This simplifies to the final relation:
% \begin{align} \label{eq:optim_m}
% m_{\text{min}} &= \texttt{ceil}\Bigg(
% \frac{d_{\text{original}} \cdot t_g}{T_2} \nonumber \\
% &\quad\cdot
% \sqrt{
% \frac{n_q}{2s}
% \cdot
% \frac{1}{\sqrt{
% \frac{[s(F_{\text{max}}-1/2)]^2}{2[\text{erf}^{-1}(2P(s)-1)]^2}
% - sF_{\text{max}}(1-F_{\text{max}})}
% }}
% \Bigg)
% \end{align}
\vspace{-1.5mm}
\begin{align} \label{eq:optim_m}
\tiny
m_{\text{min}} = \texttt{ceil}\left( \frac{d_{\text{original}} t_g}{T_2} \sqrt{ \frac{n_q}{2s {\scriptstyle\left( \frac{[s(F_{\text{max}}-1/2)]^2}{2[\text{erf}^{-1}(2P(s)-1)]^2} - sF_{\text{max}}(1-F_{\text{max}})\right)}} } \right)
\end{align}
% \vspace{-1mm}
% 
The above result is general and holds for any partitioning scheme, providing a fundamental lower bound on the number of partitions required for circuits deeper than $d_{\text{max}}$, relating hardware capabilities to algorithmic requirements. We note that, we can estimate the depths of these partitions using Equation~\ref{eq:d_max}, with the knowledge of their specific hardware parameters.

% This expression provides a general, pre-execution estimate for the minimum number of partitions needed, valid for any partitioning strategy and establishing a fundamental limit relating hardware capabilities to algorithmic requirements.

\vspace{-3mm}
\subsection{Solving Optimal Shot Allocation Problem}

% \begin{theorem}[Shot Budget Optimality]
% Given a total shot budget $S$ distributed across a quantum circuit with $m$ measurement points, the optimal allocation that minimizes total error is:
% \begin{equation}
% s_i^* = S \cdot \frac{\sqrt{\sigma_i^2}}{\sum_{j=1}^m \sqrt{\sigma_j^2}}
% \end{equation}
% where $\sigma_i^2 = F_i(1-F_i) + \sigma_{\text{noise},i}^2$ is the total variance at measurement point $i$. Any other allocation increases the estimation error.
% \end{theorem}
Given a fixed shot budget, the task of distributing shots across each partition is non-trivial, as allocating the equal number of shots would disregard the noise variances, circuit depth and qubit resources specific to a single sub-circuit, leading to significant error accumulation over multiple shots. 
To this end, we formulate finding the optimal shot allocation strategy that minimizes the total error as an optimization problem, as discussed next.

\noindent \textbf{Set Up:}\label{cond:constraint} Consider a quantum circuit with $m$ partitions $O_1, \ldots, O_m$, each requiring estimation with shots $s_1, \ldots, s_m$ for reliable execution subject to the constraint:
\vspace{-2mm}
\begin{equation}\label{eq:constraint}
\sum_{i=1}^m s_i = S
\end{equation}
% 
% \vspace{-2mm}
where $S$ is the fixed shot budget. The estimation error for observable $O_i$ measured with $s_i$ shots is \cite{preskill2018quantum} $\epsilon_i^2 = {\sigma_i^2}/{s_i}$,
% \begin{equation}
% \epsilon_i^2 = \frac{\sigma_i^2}{s_i}
% \end{equation}
where $\sigma_i^2 = F_i(1-F_i) + \sigma_{\text{noise},i}^2$ represents the total variance due to loss in quantum fidelity and hardware noise, at the measurement point $i$.
$F_i$ is obtained from per-qubit 
readout calibration data~\cite{ibmComputeResources}, and 
$\sigma^2_{\text{noise},i}$ via Eq.~\ref{eq:noise} with partition-specific 
hardware parameters.
The required number of partitions $m$ is obtained following Eq. ~\ref{eq:optim_m}.

\noindent \textbf{Objective:} Our goal is to minimize the total squared error across all measurements. Our objective function therefore is $E = \sum_{i=1}^m {\sigma_i^2}/{s_i}$.
% 
% \begin{equation}
% % E = \sum_{i=1}^m w_i \epsilon_i^2 = \sum_{i=1}^m \frac{\sigma_i^2}{s_i}
% E = \sum_{i=1}^m \frac{\sigma_i^2}{s_i}
% \end{equation}
% 
To solve the above optimization problem, we formulate the Lagrangian:
% 
% \vspace{-2mm}
% \begin{equation}
$
\mathcal{L}(s_1, \ldots, s_m, \lambda) = \sum_{i=1}^m {\sigma_i^2}/{s_i} + \lambda\left(\sum_{i=1}^m s_i - S\right)$
% \end{equation}
% 
where $\lambda$ is the Lagrange multiplier. Taking the partial derivative with respect to $s_i$ and
% setting it to zero we obtain:
% % 
% \begin{equation}
% \frac{\partial \mathcal{L}}{\partial s_i} = -\frac{\sigma_i^2}{s_i^2} + \lambda = 0
% \end{equation}
% 
% \noindent This gives:
% \begin{equation}
% \frac{\sigma_i^2}{s_i^2} = \lambda
% \end{equation}
% 
solving for $s_i$ and using the constraint from Eq. ~\ref{eq:constraint}, we have $s_i = {\sigma_i}/{\sqrt{\lambda}} \implies \sum_{i=1}^m {\sigma_i}/{\sqrt{\lambda}} = S \implies \sqrt{\lambda} = \sum_{j=1}^m \sigma_j/S$.
% \begin{equation}
% s_i^2 = \frac{\sigma_i^2}{\lambda}
% \end{equation}
% \begin{equation}\label{eq:si}
% s_i = \frac{\sigma_i}{\sqrt{\lambda}} \implies \sum_{i=1}^m \frac{\sigma_i}{\sqrt{\lambda}} = S \implies \sqrt{\lambda} = \frac{\sum_{j=1}^m \sigma_j}{S}
% \end{equation}
% 
% \noindent Using the constraint from Eq. \ref{eq:constraint}:
% \begin{equation}
% 
% \end{equation}
% \begin{equation}
% \frac{1}{\sqrt{\lambda}} \sum_{i=1}^m \sigma_i = S
% \end{equation}
% \noindent So that:
% \begin{equation}
% \sqrt{\lambda} = \frac{\sum_{j=1}^m \sigma_j}{S}
% \end{equation}
% 
% \vspace{-2mm}
Substituting back for $s_i$, we have:
%into Eq. \ref{eq:si}, we have:
\vspace{-2mm}
\begin{equation}\label{eq:optimal}
s_i^* = \frac{\sigma_i}{\sqrt{\lambda}} = \sigma_i \cdot \frac{S}{\sum_{j=1}^m \sigma_j} = S \cdot \frac{\sigma_i}{\sum_{j=1}^m \sigma_j}
\end{equation}
% \vspace{-1mm}
% 
We verify that this critical point corresponds to a local minimum by confirming the positive definiteness of the Hessian matrix.  Furthermore, the strict convexity of the objective function over the feasible region ensures the uniqueness of this solution.

% \noindent The positive definiteness of the \todo{Hessian matrix} confirms that this critical point is a minimum, and since the objective function is strictly convex over the feasible set, this solution is unique. 

\noindent\textbf{Optimality:} We prove that this allocation minimizes total error. Consider any allocation $\{s_i'\}$ satisfying $\sum_i s_i' = S$. Its total error is $E' = \sum_{i=1}^m {\sigma_i^2}/{s_i'}$.
% \begin{equation}
% E' = \sum_{i=1}^m \frac{\sigma_i^2}{s_i'}
% \end{equation}
% 
From the Cauchy-Schwarz inequality, we know $
\left(\sum_{i=1}^m s_i'\right) \cdot \left(\sum_{i=1}^m {\sigma_i^2}/{s_i'}\right) \geq \left(\sum_{i=1}^m \sigma_i\right)^2
$
% 
% \vspace{-2mm}
% \begin{equation}
% \left(\sum_{i=1}^m s_i'\right) \left(\sum_{i=1}^m \frac{\sigma_i^2}{s_i'}\right) \geq \left(\sum_{i=1}^m \sigma_i\right)^2
% \end{equation}
% 
Since $\sum_i s_i' = S$, we deduce that $E' = \sum_{i=1}^m {\sigma_i^2}/{s_i'} \geq {\left(\sum_{i=1}^m \sigma_i\right)^2}/{S}$.
% 
% \begin{equation}
% E' = \sum_{i=1}^m \frac{\sigma_i^2}{s_i'} \geq \frac{\left(\sum_{i=1}^m \sigma_i\right)^2}{S}
% \end{equation}
% 
Equality is achieved here iff $s_i' \propto \sigma_i$.
%, which is respected by our allocation rule. 
For our proposed optimal allocation (Eq. ~\ref{eq:optimal}), the error is $
E^* = \sum_{i=1}^m \frac{\sigma_i^2}{S\sigma_i/\sum_j \sigma_j} = {\left(\sum_{i=1}^m \sigma_i\right)^2}/{S}.
$
% 
% \vspace{-2mm}
% \begin{equation}
% E^* = \sum_{i=1}^m \frac{\sigma_i^2}{S\sigma_i/\sum_j \sigma_j} = \frac{\sum_j \sigma_j}{S} \sum_{i=1}^m \sigma_i = \frac{\left(\sum_{i=1}^m \sigma_i\right)^2}{S}
% \end{equation}
% 
Since $E^*$ achieves the lower bound from Cauchy-Schwarz, and $E' \geq E^*$ for any other allocation, our solution is optimal. 
Thus, for a fixed shot budget, with the knowledge of hardware characteristics, we compute the optimal shot allocation that guarantees minimizing error over all partitions.

%---------------------------------
% For a quantum computer with coherence time \(T_2\), gate time \(t_g\), and maximum fidelity \(F_{\text{max}}\), the maximum circuit depth \(d_{\text{max}}\) that maintains shot count \(s\) while preserving success probability \(P(s)\) scales as:
% \[
% d_{\text{max}} \propto \frac{T_2}{t_g} \cdot \frac{1}{\sqrt{\ln s}}
% \]
% Beyond this depth, shot requirements grow superlinearly to compensate for coherence-limited noise.
% %
% \[
% \sigma^2_{\text{noise}} = \frac{n_q}{2s}\left(2 - \frac{(t_{\text{circ}}/T_2)^2}{e^{t_{\text{circ}}/T_2}} - \cdots + \frac{p_g^2}{2}\right).
% \]
% %
\vspace{-3mm}
\section{Experimental Setup and Evaluation}
\vspace{-1mm}
\begin{figure*} [!t]
    \centering
    \includegraphics[width=1\linewidth]{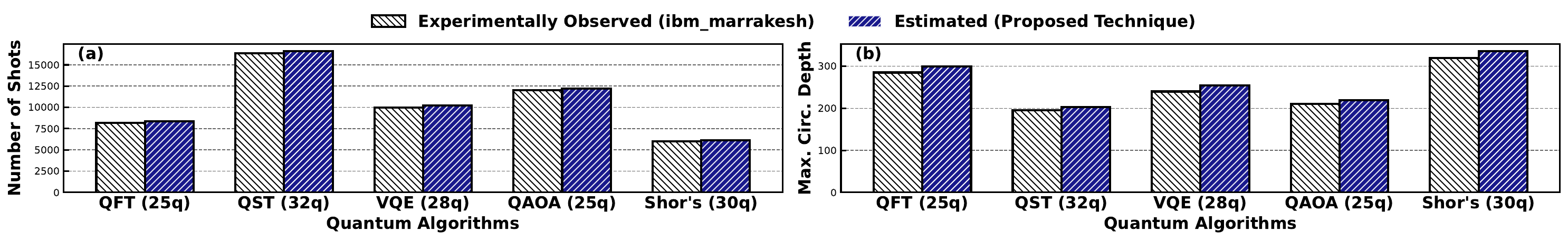}
    \vspace{-8.5mm}
    \caption{Experimental validation of proposed analytical model on IBM Marrakesh at $P(s) = 0.95$: (a) Shot estimation using Eq.~\ref{eq:shot} achieves 98.2\% accuracy on average, (b) Maximum depth estimation using Eq.~\ref{eq:d_max} achieves 95\% accuracy on average.}
    \label{fig:valid}
    \vspace{-5mm}
\end{figure*}

%\subsection{Experimental Setup}

\noindent\textbf{Quantum Algorithms Considered:} We evaluate our proposed techniques on five representative quantum algorithms: QFT for phase estimation \cite{obrien2014}, QST for state reconstruction~\cite{hou2021}, VQE for molecular energy estimation~\cite{peruzzo2014}, QAOA for combinatorial optimization~\cite{zhou2020} and Shor's Algorithm for integer factorization~\cite{martinlopez2012}. These algorithms represent diverse computational characteristics -- QFT and Shor's exhibit regular arithmetic structures, QST is measurement-dominated, VQE requires iterative classical-quantum feedback, and QAOA exploits graph locality for natural partitioning.
% , providing us good diversity for fair evaluation.

\noindent\textbf{Platform for Evaluation:}
All experiments use IBM Quantum Platform with Qiskit 2.3 (Python 3.11). Circuits are transpiled with optimization level 3 targeting native gate set (CNOT, RZ, SX, X) of IBM. Custom transpiler passes implement [[5,1,3]] error correction with syndrome extraction. Since error correction overhead is highly code-family dependent, we employ this simple stabilizer code to demonstrate indicative overhead trends.
% rather than optimal error correction performance. 
% Jobs are submitted via IBM Quantum's cloud API.
% \textcolor{blue}{We evaluate on the IBM Marrakesh (156 qubits) QPU. Hardware parameters ($F_{\text{max}}$, $T_1, T_2$, $t_\text{circ}, p_g$) are extracted from the calibration reports from IBM Quantum Platform~\cite{ibmComputeResources}}.
% %\noindent\textbf{Quantum Computers Considered:}
We evaluate our proposed technique on three IBM Quantum superconducting processors: Marrakesh (156 qubits), Torino (133 qubits) and Fez (156 qubits). Hardware parameters ($F_{\text{max}}$, $T_1, T_2$, $t_\text{circ}, p_g$) are extracted from the calibration reports from IBM Quantum Platform~\cite{ibmComputeResources}.

\vspace{-4mm}
\subsection{Verification of our Proposed Model}
\vspace{-1mm}

We verify the two key components of our proposed analytical model on real quantum hardware: quantifying the exact number of required shots using Equation~\ref{eq:shot}, and estimating the maximum feasible depth for computations on a quantum computer using Equation~\ref{eq:d_max}, given the knowledge of required success probability, hardware and algorithm parameters. We choose problem sizes such that they can be accommodated by the hardware.

Figure~\ref{fig:valid}(a) compares the number of shots required to reach $P(s) = 0.95$ between experimentally observed on \texttt{ibm\_marrakesh} and those analytically estimated by our proposed method using Equation~\ref{eq:shot}. We select $P(s) = 0.95$ as it represents a commonly accepted reliability threshold in quantum computing applications, balancing computational cost with result confidence. For all algorithms, we observe that our estimated values are very close to the values obtained experimentally. We see the most accurate results in QFT, where the error is only about 1.87\%. On the other hand, VQE results show the maximum error, of about 2.11\%. The need for classical conversion and feedback multiple times in VQE contributes to this error inherently. On average, our model is about 98.2\% accurate, validating the correctness of our theory.

Figure~\ref{fig:valid}(b) shows the comparison between the maximum computable depth for the same $P(s) = 0.95$ as earlier, recorded via experiments on \texttt{ibm\_marrakesh} and analytically estimated by our proposed method using Equation~\ref{eq:d_max}. Across multiple algorithms, we see close alignment between the experimental and estimated values, confirming that our model for estimating this maximum computable depth is indeed accurate. The maximum error (5.86\%) is seen in VQE, while QST gives the best result with about 4.41\% error. On average, our model is about 95\% accurate.
While we show extensive results on IBM Marrakesh, the performance of our model was similar on IBM Torino and IBM Fez as well, the validation error being within 2.1\% of the experimentally observed values.

% \begin{figure}
%     \centering
%     \includegraphics[width=1\linewidth]{images/dmax_ffinal_v1.pdf}
%     \caption{\todo{Validation of dmax}}
%     \label{fig:dmax}
% \end{figure}
\vspace{-3mm}
\subsection{Improvement in Number of Shots and Energy Consumption}
\begin{figure}
    \centering
    \includegraphics[width=1\linewidth]{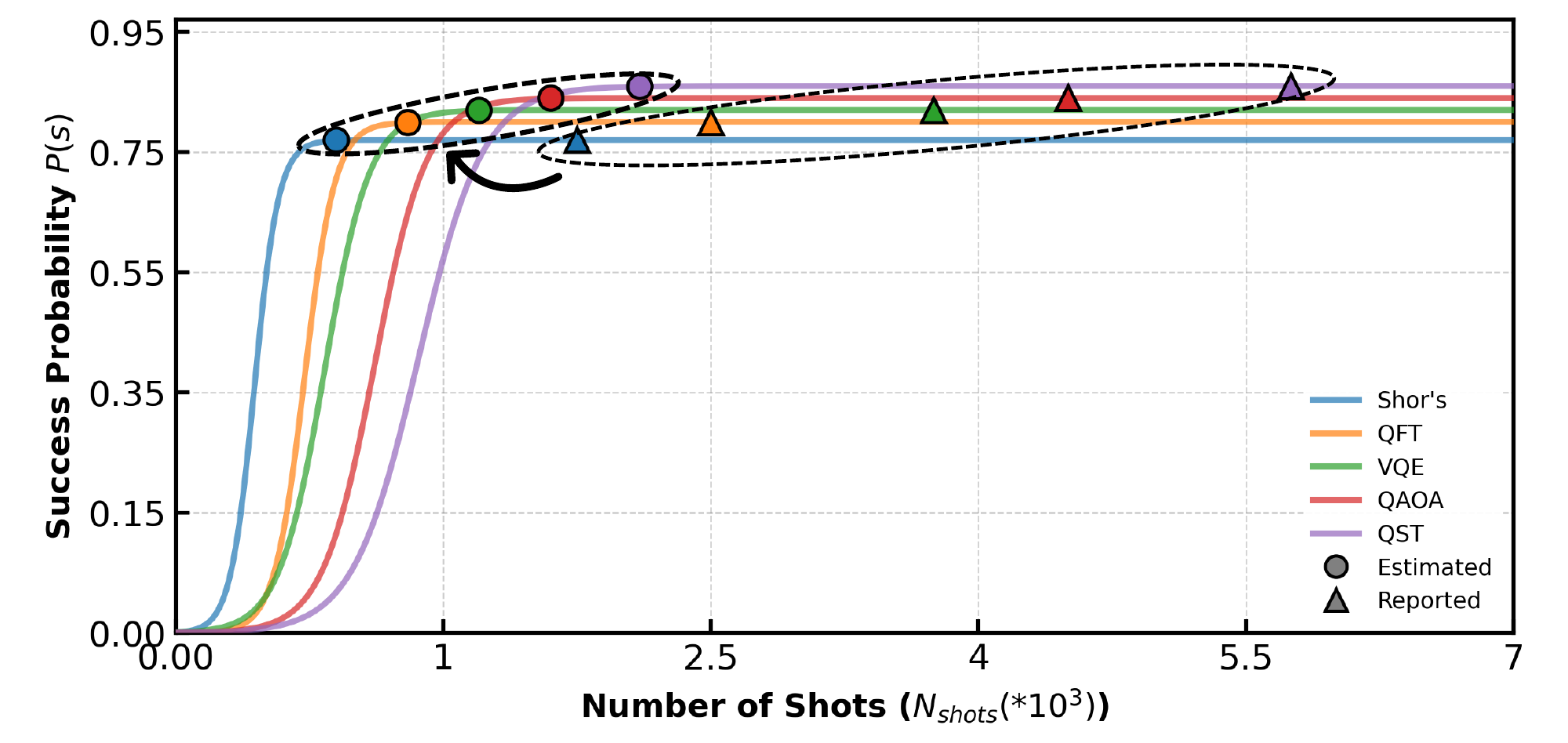}
    \vspace{-8.5mm}
    \caption{Our technique identifies optimal shot count, avoiding the wasteful over-provisioning.}
    \vspace{-4mm}
    \label{fig:probvsshots}
\end{figure}

\vspace{-1mm}
Reducing the number of shots required for quantum computations directly translates to significant practical benefits -- decreased execution time on quantum hardware, lower computational costs, and improved accessibility to resource-constrained quantum devices. 
Since quantum computers charge users based on shot count and queue time, minimizing shots without compromising result accuracy is critical for making quantum computing economically viable and scalable. Figure~\ref{fig:probvsshots} demonstrates the improvement achieved by our proposed technique compared to conventional approaches across five benchmark quantum algorithms at varying success probability thresholds ($P(s) = 0.75$ to $0.95$) \cite{zhou2020,martinlopez2012,hou2021,peruzzo2014,obrien2014}. Our method reduces shot requirements by an average of $58.6\%$ while maintaining the target success probability. These reductions are achieved through our analytical framework (Eq.~\ref{eq:shot}) that optimally balances the trade-off between measurement precision and hardware noise characteristics, eliminating the conservative over-sampling typically employed in empirical approaches. This improvement enables more efficient utilization of quantum resources and makes complex quantum algorithms more practical on current NISQ devices.~

% \textcolor{blue}{Since each shot corresponds to a single execution of the quantum circuit on the quantum processor, the total runtime and energy consumption scale approximately linearly with the number of shots executed~\cite{enriquez2023energy}. Consequently, reducing the number of shots required to achieve a target statistical accuracy directly translates to lower energy consumption. This relationship allows us to estimate the energy cost of executing different quantum algorithms by measuring the energy consumed per fixed number of shots. Figure~\ref{fig:energy_savings} shows the energy consumption per 1000 shots for different benchmark algorithms under state-of-the-art baselines and our optimal allocation strategy. Across all benchmarks, our approach consistently reduces energy consumption by allocating shots according to the variance contribution of each partition rather than distributing them uniformly, resulting in energy savings of approximately 57--62\%.}

\begin{figure}[t]
    \centering
    \includegraphics[width=1\linewidth]{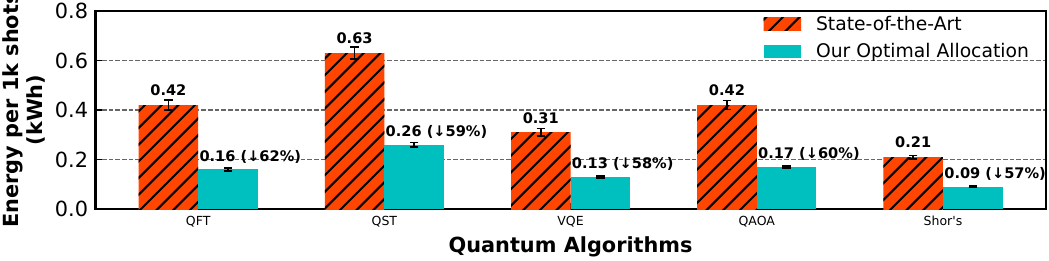}
    \vspace{-8.6mm}
   \caption{Energy consumption per 1000 shots for different quantum algorithms comparing state-of-the-art equal shot allocation with our optimal allocation strategy (20 runs, error bars indicate 95\% confidence intervals).}
   \vspace{-6mm}
    \label{fig:energy_savings}
\end{figure}

% \skm{
Figure~\ref{fig:energy_savings} compares the energy consumption per 1000 shots for different algorithms between state-of-the-art and our proposed optimal allocation strategy, for IBM Marrakesh.
The energy model is obtained from~\cite{enriquez2023energy}. This model accounts for full-stack 
power consumption, including gate operations and 
cryogenic cooling overhead.
Across all benchmarks, our approach consistently reduces energy consumption by allocating shots according to the variance contribution of each partition rather than distributing them uniformly, resulting in energy savings of up to 62\%.
We observe a similar trend in IBM Torino and IBM Fez QPUs, with energy savings of 60\% and 57.3\% on average, respectively.
% We observe a similar trend in IBM Torino and IBM Fez QPUs, with the results falling in \todo{$\pm 2\sigma$ of the values reported for IBM Marrakesh.}
%    }

% \subsection{Energy Savings via Reduced Number of Shots}

\vspace{-3.5mm}
%\subsection{Improvement in case of Distributed Quantum Computing}
\subsection{Impact on Estimation Error
}
\vspace{-1mm}
% \skm{
When a quantum circuit is partitioned into multiple subcircuits,
% due to hardware or depth constraints,
the total estimation error depends on how measurement shots are distributed across these partitions.
We, therefore, compare the estimation error obtained using our optimal shot allocation strategy against the commonly used equal-shot allocation baseline across several quantum algorithms, on IBM Marrakesh. Identifying where and how circuits should be partitioned is not the primary focus of this work. Instead, we assume a set of circuit partitions is available and study how measurement shots should be optimally allocated across these partitions to minimize the resulting estimation error.

Our shot allocation framework is independent of the circuit cutting strategy and can therefore be integrated with existing partitioning tools. We evaluate our framework using  \texttt{automatic\_cut\_finder} from the Qiskit
% circuit partitioning toolkit
~\cite{qiskit-addon-cutting}, 
which automatically determines circuit partitions that minimize sampling overhead.

\begin{figure}
    \centering
    \includegraphics[width=1\linewidth]{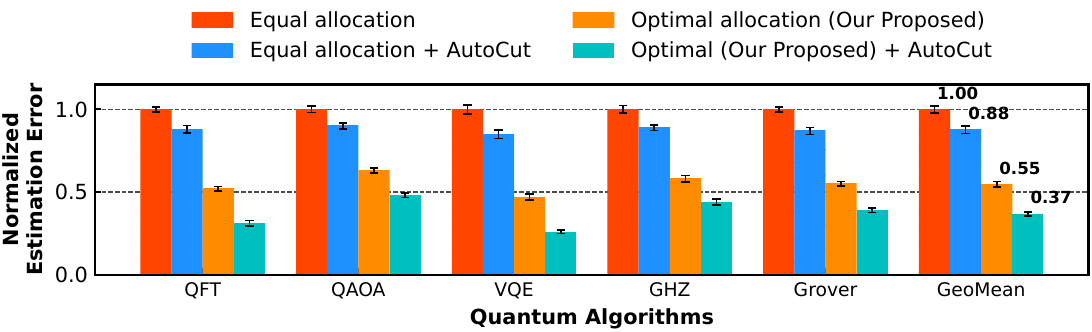}
    \vspace{-8.5mm}
    \caption{Normalized estimation error using equal allocation and our allocation, with and without automated circuit cutting (20 runs, error bars indicate 95\% confidence intervals).}
    \label{fig:autocut}
    \vspace{-5mm}
\end{figure}

Given the partitions produced by \texttt{automatic\_cut\_finder} and the corresponding shot budget, we apply our optimal allocation strategy (Eq.~\ref{eq:shot}) to distribute shots across the resulting subcircuits. We compare this variance-aware allocation against the conventional equal-shot distribution used in existing implementations. For completeness, we also include a baseline using equal shot allocation without \texttt{automatic\_cut\_finder}, which represents a naive partitioning strategy. As shown in Fig.~\ref{fig:autocut}, our approach consistently achieves lower estimation error for the same shot budget, demonstrating that our framework naturally complements automated circuit partitioning strategies.

Using the partitions generated by \texttt{automatic\_cut\_finder} and the corresponding shot budget, we apply our optimal allocation strategy to distribute shots across subcircuits. Compared to the conventional equal-shot allocation baseline, our approach reduces estimation error by an average of $45\%$ across all algorithms. Combining with automated circuit partitioning, the improvement increases to $63\%$ (Fig.~\ref{fig:autocut}). This additional gain arises because \texttt{automatic\_cut} \texttt{\_finder} produces partitions with lower sampling overhead and more balanced variance contributions, enabling our variance-aware shot allocation to utilize the shot budget more effectively. We observe a similar trend in IBM Torino and IBM Fez, with estimation error reductions of 62.5\% and 60\% on average, respectively.

\vspace{-3mm}
\section{Conclusion}
\vspace{-1mm}
Executing quantum algorithms reliably and efficiently requires determining the optimal number of measurement shots.
% while managing hardware constraints.
We derive a closed-form expression relating shots to success probability that incorporates hardware noise parameters.
% including decoherence times and gate errors.
% , with accuracy of $\textgreater 98\%$ when benchmarked on real IBM quantum hardware. Using this framework, we establish exact relationships for maximum executable circuit depth and minimum required partitions based on fundamental hardware capabilities, reaching about $95\%$ accuracy on average upon testing with IBM hardware.
% For circuits exceeding depth limits, 
Furthermore, we develop an optimal shot allocation strategy that distributes measurement resources across partitions proportionally to their noise variances. 
% Experimental validation shows our analytical model reduces shot requirements by up to 60\% on average compared to current practice while maintaining target success probability.
Our optimal allocation strategy reduces total error by 63\% on average compared to conventional equal-partition approaches across multiple quantum algorithms on IBM quantum hardware which translates to a reduction in energy consumed by about 59\%.

\vspace{-2mm}
\section*{Acknowledgments}
\vspace{-1mm}
% The authors thank the anonymous reviewers for their insightful feedback.
This work is supported by the Walmart Center for Tech Excellence at IISc (CSR Grant WMGT-23-0001).

\newpage

\balance
\bibliographystyle{ACM-Reference-Format}
\bibliography{references}

@book{nielsen2010quantum,
  title={Quantum computation and quantum information},
  author={Nielsen, Michael A and Chuang, Isaac L},
  year={2010},
  publisher={Cambridge university press}
}

@article{preskill2018quantum,
  title={Quantum computing in the NISQ era and beyond},
  author={Preskill, John},
  journal={Quantum},
  volume={2},
  pages={79},
  year={2018},
  publisher={Verein zur F{\"o}rderung des Open Access Publizierens in den Quantenwissenschaften}
}

@inproceedings{aharonov1997fault,
  title={Fault-tolerant quantum computation with constant error},
  author={Aharonov, Dorit and Ben-Or, Michael},
  booktitle={Proceedings of the twenty-ninth annual ACM symposium on Theory of computing},
  pages={176--188},
  year={1997}
}

@article{bravyi2018quantum,
  title={Quantum advantage with shallow circuits},
  author={Bravyi, Sergey and Gosset, David and K{\"o}nig, Robert},
  journal={Science},
  volume={362},
  number={6412},
  pages={308--311},
  year={2018},
  publisher={American Association for the Advancement of Science}
}

@inproceedings{shor1994algorithms,
  title={Algorithms for quantum computation: discrete logarithms and factoring},
  author={Shor, Peter W},
  booktitle={Proceedings 35th Annual Symposium on Foundations of Computer Science},
  pages={124--134},
  year={1994},
  organization={IEEE}
}

@article{Kessler2023,
  author = {Kessler, M. and Alonso, D. and Sánchez, P.},
  title = {Determination of the number of shots for Grover’s search algorithm},
  journal = {EPJ Quantum Technology},
  volume = {10},
  number = {1},
  pages = {47},
  year = {2023},
  doi = {10.1140/epjqt/s40507-023-00204-y},
  url = {https://doi.org/10.1140/epjqt/s40507-023-00204-y}
}

@article{zhu2024vqe,
  author = {Zhu, Linghua and Liang, Senwei and Yang, Chao and Li, Xiaosong},
  title = {Optimizing Shot Assignment in Variational Quantum Eigensolver Measurement},
  journal = {Journal of Chemical Theory and Computation},
  volume = {20},
  number = {6},
  pages = {2638--2648},
  year = {2024},
  doi = {10.1021/acs.jctc.3c01113},
  eprint = {2307.06504},
  archivePrefix = {arXiv}
}

@article{phalak2023qmlshots,
  author = {Phalak, Koustubh and Ghosh, Swaroop},
  title = {Shot Optimization in Quantum Machine Learning Architectures to Accelerate Training},
  journal = {IEEE Access},
  volume = {11},
  pages = {41514--41523},
  year = {2023},
  doi = {10.1109/ACCESS.2023.3270916},
  eprint = {2304.12950},
  archivePrefix = {arXiv}
}

@article{liang2024ai,
  author = {Liang, Senwei and Zhu, Linghua and Liu, Xiaolin and Yang, Chao and Li, Xiaosong},
  title = {Artificial-intelligence-driven shot reduction in quantum measurement},
  journal = {Chemical Physics Reviews},
  volume = {5},
  number = {4},
  pages = {041403},
  year = {2024},
  doi = {10.1063/5.0219663},
  eprint = {2405.02493},
  archivePrefix = {arXiv}
}

@article{mezher2025bqp,
  author = {Mezher, R. and Mills, J. and Kashefi, E. et al.},
  title = {Error Mitigation of BQP Computations using Measurement-Based Verification},
  journal = {Physical Review A},
  volume = {111},
  pages = {022602},
  year = {2025},
  doi = {10.1103/PhysRevA.111.022602},
  eprint = {2306.04351},
  archivePrefix = {arXiv}
}

@article{temme2017errormitigation,
  author = {Temme, K. and Bravyi, S. and Gambetta, J. M.},
  title = {Error Mitigation for Short-Depth Quantum Circuits},
  journal = {Physical Review Letters},
  volume = {119},
  pages = {180509},
  year = {2017},
  doi = {10.1103/PhysRevLett.119.180509}
}

@article{brandhofer2023partition,
  author = {Brandhofer, Sebastian and Polian, Ilia and Mc Kevitt, Kevin},
  title = {Optimal Partitioning of Quantum Circuits using Gate Cuts and Wire Cuts},
  journal = {arXiv preprint arXiv:2308.09567},
  year = {2023}
}

@article{martinez2019hypergraph,
  author = {Andrés-Martínez, Pablo and Heunen, Chris},
  title = {Automated distribution of quantum circuits via hypergraph partitioning},
  journal = {Physical Review A},
  volume = {100},
  pages = {032308},
  year = {2019},
  doi = {10.1103/PhysRevA.100.032308},
  eprint = {1811.10972},
  archivePrefix = {arXiv}
}

@inproceedings{tang2021cutqc,
  title={CutQC: Using Small Quantum Computers for Large Quantum Circuit Evaluations},
  author={Tang, Wei and Tomesh, Teague and Suchara, Martin and Larson, Jeffrey and Martonosi, Margaret},
  booktitle={Proceedings of the 26th ACM International Conference on Architectural Support for Programming Languages and Operating Systems},
  pages={473--486},
  year={2021}
}

@article{gidney2021factor,
  title={How to factor 2048 bit RSA integers in 8 hours using 20 million noisy qubits},
  author={Gidney, Craig and Eker{\aa}, Martin},
  journal={Quantum},
  volume={5},
  pages={433},
  year={2021},
  publisher={Verein zur F{\"o}rderung des Open Access Publizierens in den Quantenwissenschaften}
}

@misc{readout-error,
   title={QPU information},
   author={{IBM Quantum Platform}},
   year={2025},
   howpublished={\url{https://quantum.cloud.ibm.com/docs/en/guides/qpu-information}},
   note={[Accessed: 17-11-2025]}
}

@misc{ibmComputeResources,
	author = {{I}{B}{M} {Q}uantum {P}latform},
	title = {{C}ompute resources},
	howpublished = {\url{https://quantum.cloud.ibm.com/computers?order=two_q_error_best&direction=asc}},
	note = {[Accessed 17-11-2025]},
}

@article{martinlopez2012,
  title={Experimental realization of Shor's quantum factoring algorithm using qubit recycling},
  author={Martin-L{\'o}pez, Enrique and Laing, Anthony and Lawson, Thomas and Alvarez, Roberto and Zhou, Xian-Qi and O'Brien, Jeremy L},
  journal={Nature Photonics},
  volume={6},
  number={11},
  pages={773--776},
  year={2012},
  publisher={Nature Publishing Group}
}

@article{obrien2014,
  title={Quantum process tomography of a universal entangling gate implemented with Josephson phase qubits},
  author={Bialczak, Radoslaw C and Ansmann, Markus and Hofheinz, Max and Lucero, Erik and Neeley, Matthew and O’Connell, Aaron D and Sank, Daniel and Wang, Haohua and Wenner, James and Steffen, Matthias and others},
  journal={Nature Physics},
  volume={6},
  number={6},
  pages={409--413},
  year={2010},
  publisher={Nature Publishing Group UK London}
}

@article{hou2021,
  title={Benchmarking quantum state tomography on a 32-qubit superconducting quantum processor},
  author={Hou, Zhibo and Zhang, Guo-Yong and Wang, Yu and Li, Yaxin and Gu, Yifei and Li, Han and Zhu, Huangjun and Li, Guo-Long and Wang, Xiang and Song, Z and others},
  journal={PRX Quantum},
  volume={2},
  number={4},
  pages={040311},
  year={2021},
  publisher={APS}
}

@article{peruzzo2014,
  title={A variational eigenvalue solver on a photonic quantum processor},
  author={Peruzzo, Alberto and McClean, Jarrod and Shadbolt, Peter and Yung, Man-Hong and Zhou, Xiao-Qi and Love, Peter J and Aspuru-Guzik, Al{\'a}n and O'Brien, Jeremy L},
  journal={Nature Communications},
  volume={5},
  number={1},
  pages={4213},
  year={2014},
  publisher={Nature Publishing Group}
}

@article{zhou2020,
  title={Quantum approximate optimization algorithm: Performance, mechanism, and implementation on near-term devices},
  author={Zhou, Leo and Wang, Sheng-Tao and Choi, Soonwon and Pichler, Hannes and Lukin, Mikhail D},
  journal={Physical Review X},
  volume={10},
  number={2},
  pages={021067},
  year={2020},
  publisher={APS}
}

@inproceedings{enriquez2023energy,
  title={Estimating Energy-Efficiency in Quantum Optimization Algorithms},
  author={Enriquez, Rafael P. H. and others},
  booktitle={Proceedings of the Cray User Group Conference (CUG)},
  year={2023}
}

@article{barron2024provable,
  title={Provable bounds for noise-free expectation values computed from noisy samples},
  author={Barron, Samantha V and Egger, Daniel J and Pelofske, Elijah and B{\"a}rtschi, Andreas and Eidenbenz, Stephan and Lehmkuehler, Matthis and Woerner, Stefan},
  journal={Nature Computational Science},
  volume={4},
  number={11},
  pages={865--875},
  year={2024},
  publisher={Nature Publishing Group US New York}
}

@misc{qiskit-addon-cutting,
  author = {
    Agata M. Bra\'{n}czyk
    and Almudena {Carrera Vazquez}
    and Daniel J. Egger
    and Bryce Fuller
    and Julien Gacon
    and James R. Garrison
    and Jennifer R. Glick
    and Caleb Johnson
    and Saasha Joshi
    and Edwin Pednault
    and C. D. Pemmaraju
    and Pedro Rivero
    and Ibrahim Shehzad
    and Stefan Woerner
  },
  title = {{Qiskit addon: circuit cutting}},
  howpublished = {\url{https://github.com/Qiskit/qiskit-addon-cutting}},
  year = {2024},
  doi = {10.5281/zenodo.7987997}
}

@misc{seksaria2026estimatingshotsvariancenoisy,
      title={Estimating shots and variance on noisy quantum circuits}, 
      author={Manav Seksaria and Anil Prabhakar},
      year={2026},
      eprint={2501.03194},
      archivePrefix={arXiv},
      primaryClass={quant-ph},
      url={https://arxiv.org/abs/2501.03194}, 
}

\end{document}